\documentclass[a4paper,12pt]{article}

\usepackage{graphics}
\usepackage{epsfig}
\usepackage{amssymb}
\usepackage{cite}

\begin{document}
\begin{center}
{\bf Investigation of the complex dynamics and regime control in
Pierce diode with the delay feedback}\\[3mm]
 {A.E.~Hramov, and I.S.~Rempen}\\[2mm]
 {Saratov State
University,
      \it   83 Astrakhanskaja st., Saratov, 410012, Russia \\
{\small \rm
   E-mail:     aeh@cas.ssu.runnet.ru }}
\end{center}

\bigskip



\bigskip

\begin{abstract}\noindent
In this paper the dynamics of Pierce diode with overcritical
current under the influence of delay feedback is investigated. The
system without feedback demonstrates complex behaviour including
chaotic regimes. The possibility of oscillation regime control
depending on the delay feedback parameter values is shown. Also
the paper describes construction of a finite-dimensional model of
electron beam behaviour, which is based on the Galerkin
approximation by linear modes expansion. The dynamics of the model
is close to the one given by the distributed model.
\end{abstract}

\section{Introduction}

Pierce diode is one of the classical models of plasma microwave
electronics \cite{Pierce:1944, Trubetskov:2003, Granatstein:1987,
Crystal:1985, Kuhn:1990, Anfinogentov:1993, Lindsay:1995,
Kolinsky:1995, Matsumoto:1996}. This distributed model though is
rather simple, demonstrates many features of the electron beam
dynamics in different real electron devices. The model consists of
two infinite parallel plains pierced by monoenergetic electron
beam. The region between two plains is evenly filled by
neutralizing stationary ions, which density is equal to the
non-perturbed beam electron density. The only controlling
parameter is named Pierce parameter $$\alpha =\omega_{p}L/v_{0},$$
where $\omega_{p}$ is the electron beam plasma frequency, $L$ is
the distance between diode plains, $v_{0}$ is the non-perturbed
electron velocity. It was already shown that with $\alpha>\pi$
so-named Pierce instability develops in the system and the virtual
cathode is formed in the electron beam \cite{Trubetskov:2003,
Granatstein:1987}. At the same time in a narrow range of Pierce
parameter near $3\pi$ the increase of the instability is
suppressed by the non-linearity and in the electron beam the
regime without reflection realizes. In this case the system may be
described by fluid equations. It was also shown that in a narrow
range of Pierce parameter the system may represent chaotic
dynamics~\cite{Crystal:1985, Kuhn:1990, Anfinogentov:1993,
Lindsay:1995, Kolinsky:1995, Matsumoto:1996}.

Recently the possibilities of oscillation control in
finite-dimensional dynamical systems are explored in detail
\cite{Ott:1990, Pyragas:1992, Kaart:1999, Chen:1994,
Koumoou:2002}. At the same time the problem of distributed systems
regime control causes great interest. For example it may be
realized by adding to a system a controlled delay feedback
\cite{Kueny:1995, Sp-Time_St}.  The influence of the delay
feedback on the dynamics of the beam with overcritical current in
the regime of virtual cathode oscillation have been already
investigated and the influence of the feedback on the
characteristics of generation have been shown \cite{Hramov:1999}.
The problem of chaos control in Pierce diode has also attracted
the attention. In particular, the early work
\cite{Friedel:1998_ControllingChaosPierceDiode, Krahnstover:1998}
consider the possibility of stabilizating chaotic dynamics in
Pierce diode in the regime without electrons' reflection and in
the regime with virtual cathode with the help of  E.\,Ott,
C.\,Grebogy, J.\,Yorke \cite{Ott:1990} method. In our work
\cite{Koronovskii:2003} we analyse the possibility of
stabilizating of chaotic dynamics of the fluid model of Pierce
diode using the continuous delay feedback \cite{Pyragas:1992}.

But the practical realization of such schemes in microwave devices
come across several difficulties. So the investigation of  chaotic
dynamics of distributed active media with external delay feedback
causes great interest.

So it seems interesting to examine the delay feedback influence on
the dynamics of the system without virtual cathode -- the fluid
model of Pierce diode -- because it represents all the classical
regimes of a real distributed self-oscillation model. In this work
the influence of the feedback parameters to the system
characteristics is examined. The distributed chaotic system is
analysed with the help of the numerical modelling of the original
system of in partial derivative non-linear equations and with the
use of finite-dimensional model.

The structure of the work is the follows. In section 2 we discuss
the fluid model of the Pierce diode without feedback. In section 3
explored in detail the dynamics of this model with the added delay
feedback depending on the value of delay time and feedback
amplitude. The finite dimensional model of the investigated system
using the Galerkin method is constructed and its behavior under
the influence of the delay feedback is investigated in section 4.
The dynamics of finite-dimensional model is compared with the
behaviour of the electron beam in the distributed system.

\section{Explored model}

The dynamics of Pierce diode processes in fluid electronics
approximation is described by movement, continuity and Poisson
equations:
\begin{equation}\label{q1}
\frac{\partial v}{\partial t}+v\frac{\partial v}{\partial
x}=-\frac{\partial \varphi}{\partial x},
\end{equation}
\begin{equation}\label{q2}
\frac{\partial\rho}{\partial t}+ v\frac{\partial \rho}{\partial
x}+\rho\frac{\partial v}{\partial x}=0,
\end{equation}
\begin{equation}\label{q3}
\frac{\partial^2\varphi}{\partial x^2}=\alpha^2(\rho-1),
\end{equation}
where  $\varphi$ is the space charge field potential, $\rho$ is
the electron density, $v$ is the electron beam velocity.

The boundary conditions are:
\begin{equation}\label{q4}
v(0,t)=v_0,\quad
\rho(0,t)=\rho_0,\quad\varphi(0,t)=\varphi(1,t)=0.
\end{equation}

The initial conditions are taken as a small perturbation of the
space charge density near the homogeneous equilibrium state
$$
v(x)=v_{0}, \qquad \rho(x)=\rho_{0}, \qquad \varphi(x)= \varphi_0
$$
as $\rho(x,0)=\tilde\rho\sin 2\pi x$ where $\tilde\rho\ll 1$. The
equilibrium state becomes unstable then $\alpha>\pi$. In the
equations the normalized values are used.

The delay feedback is brought in by modelling the potential
difference between entrance and exit grids by the signal taken off
from the interaction space in the point $x=x_{df}$. As a control
signal the oscillations of the space charge density
$\rho(x_{df},t)$ is used. It can be interpreted as connecting a
waveguide with a delay line to the interaction space, which is
excited by the electron beam oscillations. Adding the delay
feedback into the model leads to the changes in the right boundary
conditions
$$
\varphi(1,t)=f_{df}(t)=A(\rho(x_{df},t-d)-\rho_0).
$$
Here $A$ is the delay feedback coefficient, characterizing the
part of the oscillation power  branched to the feedback delay
line, $d$ is the delay time value. Assuming that the development
of the processes in our system begins at $t=0$ and when $t<0$ the
space charge density is non-perturbed $\rho(x,t)=\rho_0$, the
initial distribution of the delay feedback function is written as
$$
f_{df}(t)|_{t\in[-d,0]}=0.
$$

We have found out that the point of connection $x_{df}$ does not
influence upon the dynamics of the system. In this work value
$x_{df}$ is fixed as $x_{df}=0.2$.

Numerical solution of the equations (1) and (2) is found using
explicit scheme with differences against flow. Poisson equation
(3) is integrated using error vector propagation (EVP) method
\cite{Roache:1972}.

\section{Delay feedback influence on the nonlinear dynamics of electron beam}

In Pierce diode without delay feedback when $\alpha$ decreases
from $2.88\pi$ to $2.86\pi$ the behaviour of the electron beam
changes from regular via period doubling cascade to weakly chaotic
with neatly expressed time scale.  With the further decrease of
Pierce parameter the chaotic oscillations of the beam complicate
essentially, the time scale disappears and spectral distribution
complicates. We call this two types of chaotic behaviour ribbon
and spiral chaos. All the results described in this paper have
been derived for $\alpha=2.86\pi$, i.e. a system without delay
feedback must represent the ''spiral chaos'' oscillations. As
quantitative characteristics of oscillation regime correlation
dimension $D$ \cite{Grassberger:1983} and highest Lyapunov index
$\lambda$ \cite{Wolf:1989} for the restored attractor are taken.
This values do not change in the different points of interaction
space $D=2.18\div0.01$, $\lambda=0.16\div0.04$. In figure 1 the
different regimes are represented on the $A$--$d$ parameter plane
with $\alpha=2.86\pi$.

\begin{figure}[h]
\centerline{\includegraphics[width=8cm]{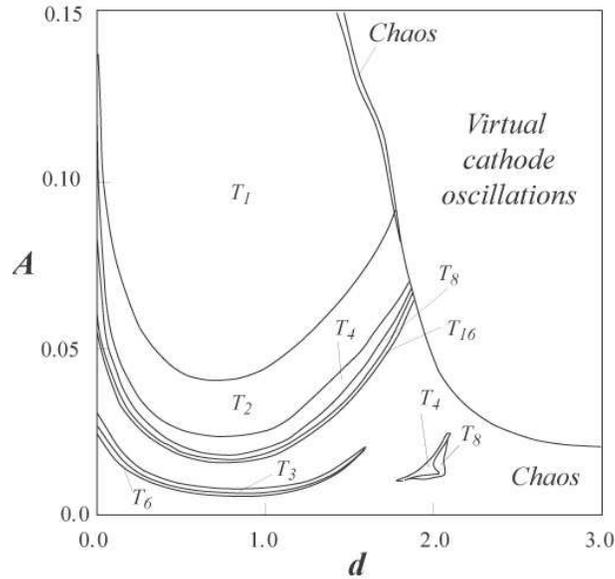}}
\caption{Oscillation regimes reproduced on parameter plane ($A$,
$d$) ($\alpha=2.86\pi$, $x_{df}=0.2$)}
\end{figure}

For comparison the non-dimensional characteristic time of
oscillations in the electron beam $\tau=4.06$. The areas of
$n$-periodical oscillations on the parameter plane are marked as
$T_{n}$ When $A\ll1$ the system demonstrates chaotic oscillations
identical to those without feedback.

\begin{figure*}[t]
\centerline{
\includegraphics[width=14cm]{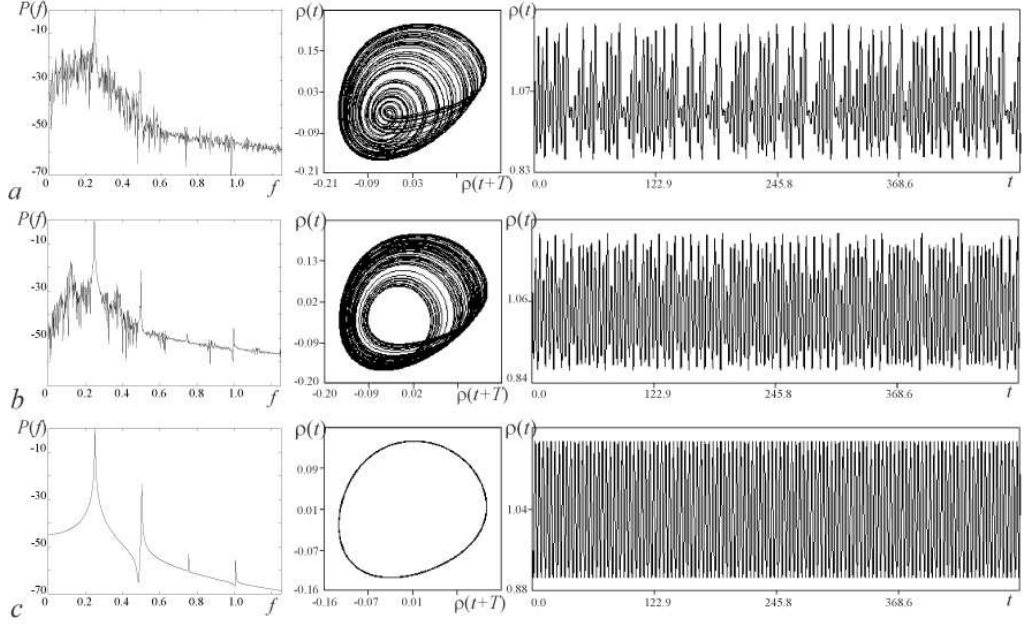}}
\caption{Phase portraits, spectrums and time series of electron
beam oscillations for the cases of ``spiral'' chaos regime (a),
weakly developed chaos (b) and regular (period 1) (c) regimes}
\end{figure*}

In figure 2 the phase-plane portraits, spectrums and time series
for the ''spiral'' chaos regime (a), weakly developed chaos (b)
and regular (period 1) oscillations (c) are represented. The
phase-plane portraits are reconstructed by Takens delay method
\cite{Takens} from the time series characterizing the oscillation
of the space charge density in the fixed point of the distributed
system. Though system behaviour (a) is rather complex, in the
frequency spectrum one can see the base frequency
$f_{0}=1/\tau=0.25$ and its second harmony $2f_{0}$. Analyzing the
system attractor in phase-plane space one can see that near the
instable state $\rho=\rho_{0}$ there is a loop on which the motion
of the phase point become slower. The other space of attractor is
tightly filled by spiral phase pathes. When $A$ increases the
behaviour of the system may be different depending on the value of
delay time $d$. If $d>\tau/2$ the complexity of electron beam
dynamics increases with the increasing of $A$. The frequency
spectrum and the phase-plane portrait become more complex too, the
oscillation amplitude enlarges. Further enlargement of $A$ leads
to essentially different dynamics of the system. The oscillation
amplitude sharply increases and then in the electron beam the
reflected electrons appear. In the system appears the virtual
cathode. The electron beam behaviour is determined by two
mechanisms - the Pierce instability and its limitation by the
nonlinearity. The delay feedback with parameter values
$d>1.8\div2$ è $A>0.02$ destroys the limitation mechanism and
leads to the increasing of instability and further to virtual
cathode forming. In this case the fluid model becomes incorrect
because equations (1), (2) describes the processes in the electron
beam only without overtaking or reflection. The threshold value
$A_{VC}$ depends on the value of $d$.
 In the case $d<\tau/2$ the complexity of the electron beam oscillation
decreases with the increase of $A$. The noise base diminishes,
with the further enlargement of $A$ on the bifurcation map one can
see periodic gaps. When $A>0.03$ a transition from the chaotic to
periodical dynamics via reverse doubling period cascade have
place.  The oscillation amplitude decreases and approach to
non-perturbed value $\rho_{0}$. In a wide range of feedback
parameters it is possible to suppress the chaotic dynamics of the
electron beam and to establish the regular one. In figure 1 the
areas of cycles 1-16 period and transitions between different
regimes depending on the changes of delay feedback parameters are
shown.

Now some words about the physical processes in the electron beam.
As the explorations of the electron waves propagation show,
dynamics of electron beam is mainly determined by the distance
between the current state of the system and the homogeneous
equilibrium state $\rho(x)=\rho_0$, $v(x)=v_0$,
$\varphi(x)=\varphi_0$. The distance between the current state of
the system and the homogeneous equilibrium state can be determined
as
\begin{equation}\label{q5}
S(t)=\left(\int\limits_0^1(\rho(x,t)-\rho_0)^2+
 (\varphi(x,t)-\varphi_0)^2+(v(x,t)-v_0)^2\, dx\right)^{{1}/{2}}.
\end{equation}

\begin{figure}[h!]
     \leavevmode
\centering
\includegraphics[width=8cm]{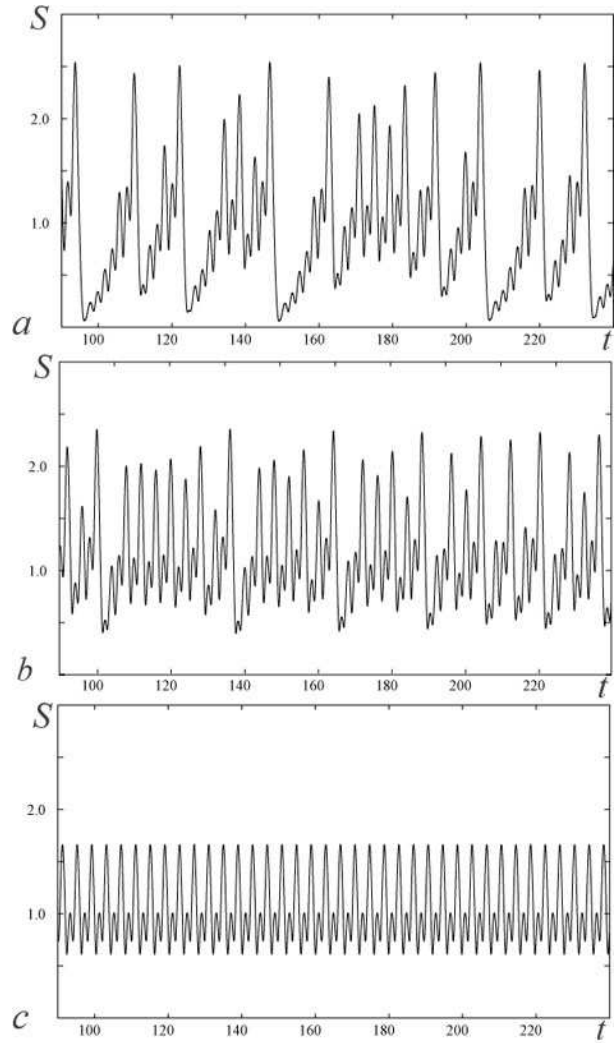}
\caption{The distance between the current state of the system and
the homogeneous equilibrium state depending on time for the cases
of ''spiral'' chaos regime (a), weakly developed chaos (b) and
regular (period 1) (c) regimes}
\end{figure}

Time-dependent changes of this value illustrates figure 3 for the
cases of spiral chaos regime (a), weakly developed chaos (b) and
regular (period~1) oscillations~(c).

In the first case the system in some time comes very close to the
homogeneous equilibrium state $S \sim 0$ and the oscillation
amplitude is very small. Then the mechanism of instability
activates and the amplitude of the oscillations increases until
it's limited by the nonlinearity. Then the process repeats, but
each time $S$ and the space distributions of the values near the
equilibrium state are different, so the development of instability
begins from another conditions and the dynamics of the system is
irregular. The dynamics of the system can also be examined by
considering non-linear energetic functionals \cite{Yagata:1987}:
\begin{equation}\label{q6}
W_{k}=\frac{1}{2}{\int\limits_0^1\rho v^2\, dx}-\frac{1}{2},\quad
W_{p}=\frac{1}{2}{\int\limits_0^1\rho \varphi\, dx}
\end{equation}

This functions describe energy transitions between kinetic energy
of the beam movement and potential energy of the space charge
field. One can see that for the chaotic regimes the maximums of
the functionals are larger than those for the regular processes,
because of the larger degree of non-linearity  in the chaotic
regime.

\begin{figure}[h!]
     \leavevmode
\centering
\includegraphics[width=8cm]{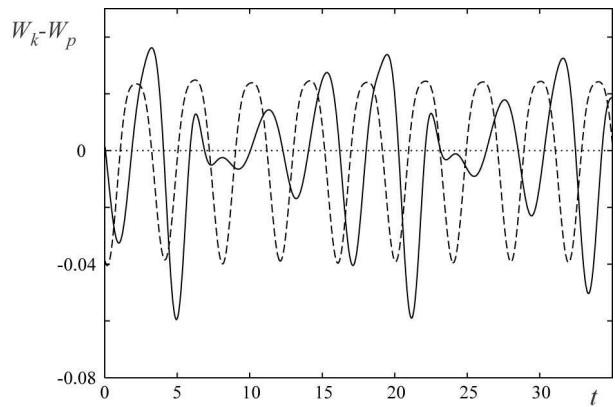}
\caption{Difference between kinetic energy and potential energy of
the beam movement $\Delta W=W_{k}-W_{p}$ for the cases of
''spiral'' chaos regime (solid line) and regular regime (dotted
line) }
\end{figure}

Figure 4 illustrates the time-dependence of the function $\Delta
W=W_{k}-W_{p}$ characterizing the energy transition processes in
the electron beam. In the periodic regime (the dotted line) the
energy transitions are regular. The maximums of the functions
shows the accumulation of the charge in the interaction space.
    In the chaotic regime (the solid line) in some moments the energy
 difference $\Delta W=0$. The system nears to the unstable equilibrium
state. Then the wave movement energy increases again abruptly.
Further the perturbation energy decreases and the system
protractedly is situated near the equilibrium state.  It can be
physically explained as follows: in the electron beam the
stationary wave of electron density appears. Its amplitude
increases abruptly near the exit grid. Then the electromagnetic
field of this clot brakes the following electrons and the result
is that much more electrons leave the interaction gap then enter
it. The discussed values approach the equilibrium state and then
the process repeats from the new starting point. The feedback
destroys this mechanism. When the delay time $d<\tau/2$ the system
cannot approach the equilibrium state because the feedback signal
extenuates the stored charge in the exit region and promotes the
acceleration of the beam in the instant time when $\Delta W$ is
maximal. And vice versa when $d<\tau/2$ the feedback signal leads
to the increasing of oscillation amplitude, the system dynamics
become more complicate and with sufficiently large $A$ the virtual
cathode appears.

\section{Finite-dimensional model of electron beam
dynamics}

In \cite{Trubetskov:2003} the method of constructing a
finite-dimensional model based on the Galerkin approximation by
linear modes expansion is described. It has been shown by
\cite{Godfrey:1987} that in the range of Pierce parameter
variation $\alpha \in (2\pi,3\pi)$ in the system excites infinite
number of modes which can be determined from the dispersion
equation

$$
\left\{\exp[{j\alpha\varpi}]\left[(\varpi^2+1)\sin\alpha+2j\varpi\cos\alpha\right]
+\alpha\varpi^4-\right.
$$
\begin{equation}\label{q7}
\left.\qquad\qquad-\alpha\varpi^2-2j\varpi\right\}{(\varpi^2-1)^{-2}}=0,
\end{equation}
where $\varpi=\omega/\omega_{p}$. In the case of Pierce diode
system the modes were determined by Kuhn (1986). It have been
shown that among the excited modes only three were damping rather
slowly and containing the most part of the system energy.  So for
the description of the system dynamics it is enough to take into
account only this three modes. For the different Pierce parameter
values corresponding to different dynamical regimes the space
distributions vary weakly, so we can suppose they are independent
of $\alpha$. The initial basis for the finite-dimensional
approximation is taken as
$$
v=\sum\limits_{i=1}^3V_{i}(x)a_{i}(t),
$$
\begin{equation}\label{q8}
\rho=\sum\limits_{i=1}^3R_i(x)a_i(t),
\end{equation}
$$
\varphi=\sum\limits_{i=1}^3\Phi_i(x)a_i(t),
$$
where $R_{i}$, $V_{i}$, $\Phi_{i}$ are the space distributions
modes for the $\lambda_{i}$, $\alpha_{i}$ are the modes
amplitudes. Substituting the trial solution (8) into system
(1)-(3), written for weakly perturbed values, we derive the
nullity vector $\vec\Psi=(\Psi_1,\Psi_2,\Psi_3)$, which components
can be written as
$$
\Psi_1=\dot{a}_1R_1+\dot{a}_2R_2+\dot{a}_3R_3+a_1
({R_{1x}+V_{1x}})+
$$
$$
+
 a_2({R_{2x}+V_{2x}})+a_3({R_{3x}+V_{3x}})+
$$
$$
+a_1^2(R_1V_1)_x+a_2^2(R_2V_2)_x+a_3^2(R_3V_3)_x+
$$
$$
+a_1a_2(R_1V_2+R_2V_1)_x+a_1a_3(R_1V_3+R_3V_1)_x+
$$
\begin{equation}\label{q9}
+a_2a_3(R_2V_3+R_3V_2)_x,
\end{equation}
$$
\Psi_2=\dot{a}_1V_1+\dot{a}_2V_2+\dot{a}_3V_3+a_1({\Phi_{1x}+V_{1x}})+
$$
$$
 +a_2({\Phi_{2x}+V_{2x}})+a_3({\Phi_{3x}+V_{3x}})+
$$
$$
 +a_1^2V_1V_{1x}+a_2^2V_2V_{2x}+a_3^2V_3V_{3x}+
$$
$$
+a_1a_2(V_1V_{2x}+V_2V_{1x})+a_1a_3(V_1V_{3x}+V_3V_{1x})+
$$
\begin{equation}\label{q10}
+a_2a_3(V_2V_{3x}+V_3V_{2x}),
\end{equation}
$$
\Psi_3=a_1(\Phi_{1xx}+\alpha^2R_1)+a_2(\Phi_{2xx}+\alpha^2R_2)+
$$
\begin{equation}\label{q11} +a_3(\Phi_{3xx}+\alpha^2R_3),
\end{equation}
where $(\cdot)_x=\frac{\partial}{\partial x}(\cdot)$.

The internal product of functions is defined as
\begin{equation}\label{q12}
({f\times g})=\int\limits_0^1fg\,dx.
\end{equation}

Using Galerkin method we can find the unknown coefficients $a_{i}$
from the matrix equation
\begin{equation}\label{q13}
\left(\left(
  \begin{array}{ccc}
    R_1 & V_1 & \Phi_1 \\
    R_2 & V_2 & \Phi_2 \\
    R_3 & V_3 & \Phi_3
  \end{array}\right)
\times
 \left(
  \begin{array}{c}
    \Psi_1 \\
    \Psi_2 \\
    \Psi_3
  \end{array}\right)\right)=0.
\end{equation}
Carrying out elementary transformation and taking into account the
equations (9)--(11), we derive the matrix equation for the
coefficients $a_{i}$:
\begin{equation}\label{q14}
{\bf M}\dot{{\bf A}}+{\bf B}{\bf A}+ {\bf D}=0,
\end{equation}
where vector ${\bf A}$ is composed from coefficients $a_{i}$. The
elements of matrixes ${\bf M}$ and ${\bf B}$ are derived as
$$
m_{i,j}=(R_{j}\times R_i)+(V_j\times R_i),
$$
$$
b_{i,j}=\left((R_{ix}+V_{ix})\times
R_j\right)+\left((V_{ix}+\Phi_{ix})\times V_j\right)+
$$
$$
+((\Phi_{ixx}+\alpha^2R_i)\times \Phi_j).
$$
Matrix element ${\bf D}$ is derived from formula
$$
d_{i}=\sum\limits_{k}a_k^2 \left[{(R_kV_k)_xR_i
+V_kV_{kx}V_i}\right]+
$$
$$
+\sum\limits_{k}\sum\limits_{l,~l\neq k}a_ka_l\left[
(R_kV_l+R_lV_k)_xR_i+\right.
$$
$$
\left. +(V_kV_{lx}+V_lV_{kx})V_i \right].
$$

Resolving the equations (14) relatively $\dot{a}_{i}$, we derive
the explicit equations:
$$
\dot{a}_1(t)=l_{11}a_1+l_{12}a_2+l_{13}a_3+l_{14}a_1^2
+l_{15}a_2^2+
$$
\begin{equation}\label{q15}
+l_{16}a_3^2+l_{17}a_2a_3+l_{18}a_1a_1+l_{19}a_1a_3,
\end{equation}
$$
\dot{a}_2(t)=l_{21}a_1+l_{22}a_2+l_{23}a_3+l_{24}a_1^2
+l_{25}a_2^2+
$$
\begin{equation}\label{q16}
+l_{26}a_3^2+l_{27}a_2a_3+l_{28}a_1a_1+l_{29}a_1a_3,
\end{equation}
$$
\dot{a}_3(t)=l_{31}a_1+l_{32}a_2+l_{33}a_3+l_{34}a_1^2
+l_{35}a_2^2+
$$
\begin{equation}\label{q17}
+l_{36}a_3^2+l_{37}a_2a_3+l_{38}a_1a_1+l_{39}a_1a_3.
\end{equation}
The coefficients $\l_{ij}$ are derived from numerical solution of
the implicit equations and are independent of Pierce parameter
variations. The non-linearities of the system are quadratic and
appears because of the kinematic non-linearities and those
presenting in continuity equation. The 1 and 2 modes are excited
by the instability negative dissipation and its energy is
inherited into the 3rd mode. As the numerical analysis shows the
finite-dimensional demonstrates the same types of behaviour as the
distributed one. With the decrease of $\alpha$ the system dynamics
becomes more complex and further transition to chaos via period
doubling cascade takes place. The comparison of bifurcation
diagrams for the distributed model and the finite-dimensional
model is made in figure 5.

\begin{figure}[h!]
     \leavevmode
\centering
\includegraphics[width=8cm]{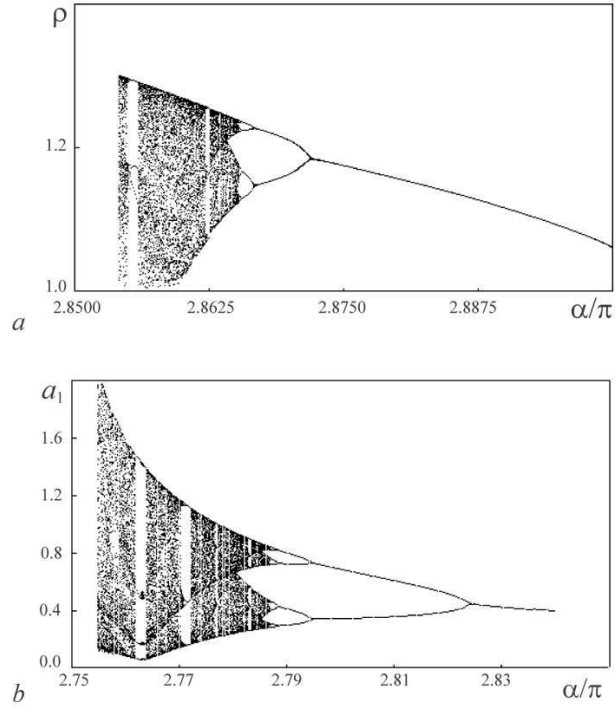}
\caption{Bifurcation diagrams for the distributed model (a) and
the finite-dimensional model (b)}
\end{figure}

Also in the system two variants of the chaotic regime similar to
those in the distributed model are observed -- the ''spiral''
chaos and the ''band'' chaos. The delay feedback is brought in by
adding into the right part of the equations (7)-(9) the signal
$F_{df}(t-d)$ which is formed as:
\begin{equation}\label{q18}
F_{df}(t)=A[a_1(t)+a_2(t)+a_3(t)]
\end{equation}
In figure 6 the bifurcation map in the parameters $A$ -- $d$ for
the finite-dimensional model is presented. The Pierce parameter
$\alpha=2.774\pi$. With this parameter value in the system without
feedback the ''spiral'' chaotic oscillations are observed. In the
map we can see that with the increase of feedback signal amplitude
and $d<\tau/2$ (where $\tau \approx 3.15=1/f_0$, $f_0$ - the base
spectrum frequency) the chaotic dynamics is suppressed and regular
regime is installed.

\begin{figure}[h!]
     \leavevmode
\centering
\includegraphics[width=8cm]{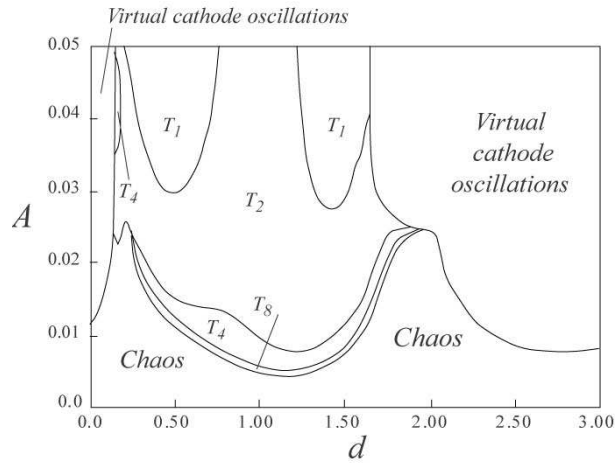}
\caption{$A$--$d$ parameter plane for the finite-dimensional model
with delay feedback, $\alpha=2.86\pi$}
\end{figure}

When $d>\tau/2$ the oscillation amplitude sharply increases which
is equal to system transition to virtual cathode forming regime,
where the finite-dimensional equations become incorrect. Comparing
figure 1 and figure 4 one can see that the finite-dimensional
model gives a very good description of the processes taking place
in the distributed system.

\section{Conclusion}

In our work the delay feedback influence on the electron beam
dynamics in hydrodynamical and finite-dimensional models of Pierce
diode is investigated. It is shown that with some feedback
parameters' values the chaotic dynamics of the electron beam is
suppressed and periodical regimes of different types may be
installed. Physically it is connected with the changing of
conditions of electron waves propagation. The practical interest
for this phenomenon is caused by the ability of eliminating the
undesirable parasitical and chaotical oscillations in some real
systems where Pierce instability may appear (for example, in
electron  guns, beams of charged particles, etc.).

\section*{Acknowledgements}

We are thankful to Corresponding Member of Russian Academy of
Sciences, Prof. D.I.\,Trubetskov for the fruitful discussion of
our work.

The work is supported by Russian Basic Research Fund (grant No
02--02--16351) and grant REC--006 of U.S. Civilian Research \&
Development Foundation for the Independent States of the Former
Soviet Union (CRDF)).

\end{document}